\documentclass[final, runningheads]{llncs}
\usepackage[T1]{fontenc}
\usepackage{graphicx}
\usepackage{amsmath}
\newcommand{\itadata}{\footnotesize \textsl{ITADATA2024: The 3$^{\text{rd}}$ Italian Conference on Big Data and Data Science}}
\usepackage{fancyhdr}
\fancypagestyle{empty}{\fancyhf{}\fancyhead[C]{\itadata}
  \fancyhead[R]{\includegraphics[width=.05\textwidth]{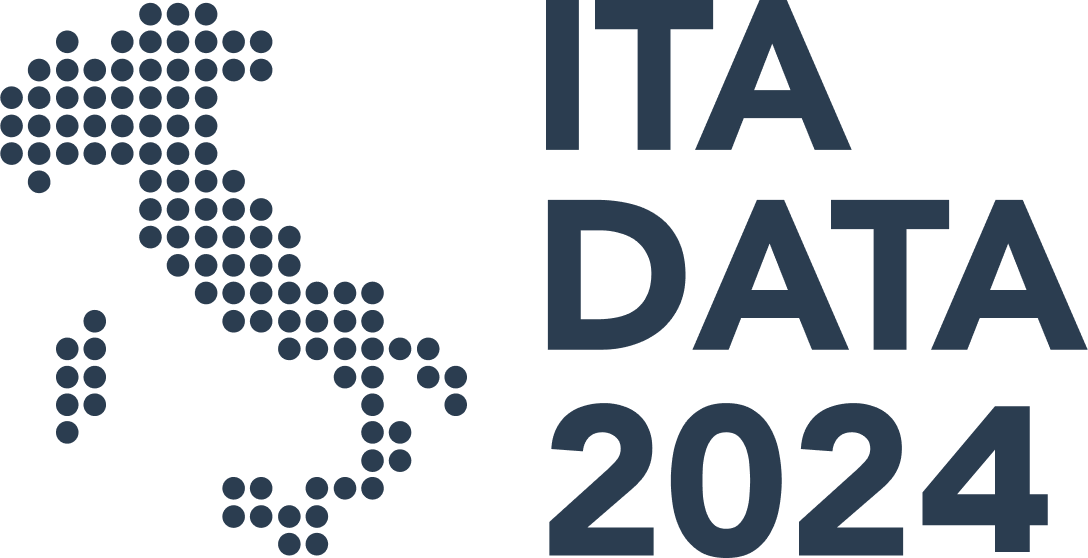}}}

\makeatletter
\begin{document}
\title{Financial Twin Chain, a platform to support financial sustainability in supply chains}
\author{
Giuseppe Galante \inst{1}  \and
Christiancarmine Esposito\inst{1}\and
Pietro Catalano\inst{2} \and
Salvatore Moscariello\inst{2} \and
Pasquale Perillo\inst{2} \and
Pietro D'Ambrosio\inst{2} \and
Angelo	Ciaramella\inst{3} \and
Michele Di Capua\inst{4}
}

\authorrunning{F. Author et al.}
\institute{Università degli Studi di Salerno - Via Giovanni Paolo II, 132 - 84084 Fisciano (SA)  \and
LinearIT spa. Via Giovanni Severano, 28, 00161 Rome, Italy \and
DIST – Università di Napoli “Parthenope”, Naples, 80143, Italy. \and
US srl, via Porzio, Centro Direzionale di Napoli Isola G2, Naples, 80143, Italy.
}

\maketitle              \begin{abstract}
The financial sustainability of a generic supply chain is a complex problem, which can be addressed through detailed monitoring of financial operations deriving from stakeholder interrelationships and consequent analysis of these financial data to compute the relative economic indicators. This allows the identification of specific fintech tools that can be selected to mitigate financial risks. The intention is to retrieve the financial transactions and private information of stakeholders involved in the supply chain to construct a knowledge base and a digital twin representation that can be used to visualize, analyze, and mitigate the issues associated with the financial sustainability of the chain. We propose a software platform that employs key enabling technologies, including AI, blockchain, knowledge graph, and others, opportunely coordinated to address the financial sustainability problem affecting single stakeholders and the entire supply chain. This platform allows for the involvement of external entities that can help stakeholders or the whole supply chain to solve financial sustainability problems through economic interventions. Moreover, introducing these entities enables stakeholders less well-positioned in the market to access financial services offered by credit institutions, utilising the supply chain's internal information as evidence of its reliability.
To validate the proposed idea, a case study will be presented analyzing the financial instrument of securitization.

\end{abstract}
\section{Introduction}
The financial sustainability of a supply chain (namely Supply Chain Finance (SCF)\cite{Ioannou}) is a critical issue where each stakeholder has well-defined and separated financial and functional autonomy organized and supervised to create value (products or services). 
SCF is a micro-finance concept defined as the use of financial instruments, practices, and technologies for optimizing the management of the working capital and liquidity tied up in supply chain processes between collaborating business partners and, more generally, in a generic industrial configuration where each division have well defined and separated financial and functional autonomy, that are organized and supervised, to create value (products or services). The decentralized nature of a supply chain (SC) makes its financial sustainability a challenging aspect to realize, and there are no easy ways to deal with it. For this reason, this paper proposes a platform for monitoring a supply chain and the consequent support to the introduction of ways to correct possible losses or inefficiencies.\\
The proposal employs a subset of essential and creative solutions in the landscape of computer science, including the design of a digital twin system to reflect the financial health of the SC, the utilization of a blockchain network and smart contracts to address the aspects of certification and traceability of financial operations, the continuous monitoring  of supply chain’s financial sustainability-related aspects to predict (through AI services \cite{Fengyan}) problematics events, and the proposal of tailored fintech services that can be used to mitigate the relative alert scenarios. Overall, the financial dimension in managing the SC processes allows us to reinforce the new level of trust we want to achieve by applying the proposed service model. This trust is not limited to the economic actors within the supply chain but extends to the end consumer.\\
This paper is inspired by the ongoing activities regarding a current industrial research work and aims to present a reporting of the analyses carried out within the SCF that define the target challenges to be addressed, as well as a design of the aforementioned platform that can facilitate the collaborative utilization of the listed technologies. The platform is designed to be adaptable to different supply chain contexts, both new and pre-existing, and approach to different analysis models and supporting services.\\
The remainder of the paper is organised as follows. Section \ref{review} analyses the existing literature on the topic, while Section \ref{approach} describes the proposed methodology. Section \ref{risk} highlights the financial risks, and Section \ref{case} describes the validation case. The platform is presented in Section \ref{platform}. Final remarks and future work are contained in Section \ref{end}.

\section{Literature Review}
\label{review}

This section aims to detail the analysis we have performed on the current state of fintech tools and techniques, focusing on identifying potential applications of emerging technologies to solve sustainability challenges in supply chain environments.  
The current challenges primarily relate to financial management within the SC. It is vital to maintain stable cash flows to ensure the efficient conduct of daily operations and to meet demand levels
\cite{Olan}.  Financial management is complex mainly due to the diversity of actors involved, the heterogeneity of accounting processes, and the necessity to coordinate financial activities among stakeholders.  The lack of tools and methodologies to proactively assess and manage financial risk can impede a company's capacity to guarantee its financial activities' stability and sustainability and capitalize on growth and development opportunities \cite{Tseng}. SC financial transactions are frequently fragmented and isolated, particularly in small and medium-sized enterprises. This lack of traceability and transparency makes it more challenging for stakeholders to ensure the integrity and accuracy of economic transactions, which can result in issues such as late payments and suspended sales. \cite{Kaur}\cite{Zhu}.\\
Several SCF instruments have been designed as solutions to address the issues mentioned above. For example, inventory financing \cite{Ioannou} addresses the lack of liquidity, enabling companies in the supply chain to self-finance. Similarly, factoring or reverse factoring \cite{Ioannou} enables companies to seek financing from financial institutions.
In addition, research has found a significant relationship between information technology and supply chain management. Among the most impactful technologies is AI, which has been identified as a pivotal innovation in enhancing SC efficiency and effectiveness \cite{Olan}. Big data analysis has been used to evaluate the financial sustainability of supply chains at the regional level, enabling a detailed assessment of the performance of SCFs in different geographic areas \cite{Tseng}. Blockchain technology also has a significant impact on the SCF. Its introduction into fintech services can be used to create value within the supply chain by providing transparency, traceability, and security to financial transactions occurring in the supply chain. \cite{Wang}\cite{Renduchintala}. The features of decentralization, encryption, and the consensus mechanism, in addition to transparency and immutability of data, can enhance trust among parties involved in SC, enabling them to engage in financial transactions without requiring a centralized authority to verify them. Improved visibility and risk assessment can help overcome cash flow challenges by opening new avenues for access to credit and improving cash flow management within the supply chain.   Furthermore, smart contracts facilitate trading efficiency, allowing financial transactions to be executed automatically under certain conditions or clauses \cite{Renduchintala}.\\
The research findings indicate that the effective management of financial risk in the supply chain is essential to ensuring the business continuity and financial sustainability of the companies involved. To achieve this, enhanced communication among members is needed, which would also promote further demand for SCF services.    

\section{The methodological approach}
\label{approach}

The holistic approach of the proposed work provides an augmented visualization of the entire process and assets derived from the big data produced throughout the supply chain lifecycle. Financial transactions certified on the blockchain \cite{Park} and private stakeholders’ information enhance chain knowledge, enabling accurate analysis for financial sustainability.\\
While we focus on SCF, it is logical to consider a more comprehensive model (using, for example, AI functionalities \cite{Trong}) that utilizes the whole data from a complex supply chain network, including IoT, product line data, and so forth. This extension can build a digital twin for additional analyses, monitoring, and suggesting solutions for supply chain management (SCM) \cite{Swaminathan} challenges, ethical, and environmental impacts.\\
The proposed system is designed to achieve a high degree of expandability, allowing the introduction of a set of monitoring tools (for example, new AI services) and financial technology services within a well-defined “pluggable” environment model, to extend the base functionalities of the system. To achieve this, we have designed an architectural solution that manages the large volumes of data generated throughout the supply chain lifecycle. Another added value is the flexible design of the knowledge base, which employs a semantic knowledge graph to collect and combine data from different sources, enhancing the pre-existing information with semantic enrichment and enabling the generation of inferences and new wisdom about the facts of the supply chain.
Finally, the system is open to external entities, who can invest by utilizing the platform’s fintech tools to inject new liquidity into the financial circuit.

\section{Financial risk and services}
\label{risk}
In the context of business risk, insolvency of a company is defined as the inability to meet its financial obligations to creditors as debts become due \cite{insolvenza}. This is in contrast to corporate solvency, which refers to the concrete ability of the company to meet its long-term debts and financial obligations \cite{solvenza}. 
Both are key factors in evaluating financial institutions' creditworthiness and determining credit accessibility for companies within the supply chain. It is important to note, however, that to assess a stakeholder's financial solvency and financial risk, it is necessary to monitor their behaviour both inside and outside the supply chain. 
However, some tax information can only be viewed outside the supply chain, mainly concerning compliance with tax regulations and transparency towards the authorities.\\
Table \ref{index} provides an analytical overview of the financial indexes and ratios that can influence the definition of corporate risk.
Table \ref{services} lists some of the solutions \cite{Ioannou}, indicating which of the previously mentioned indexes are most pertinent to the service in question.

\begin{table}[t]
\centering
\scriptsize
\caption{List of financial indexes}
\label{index}
\begin{tabular}{|l|l|l|l|}
\hline
Index & Description & Good values & Alert values \\ \hline
Debt index & \begin{tabular}[c]{@{}l@{}}The debt ratio is an indicator that allows \\ to measure the situation in which a \\ company finds itself to its debts. \cite{indebitamento}\end{tabular} & \begin{tabular}[c]{@{}l@{}}Ratios less than \\ 0.5 or 50\%\end{tabular} & \begin{tabular}[c]{@{}l@{}}Greater than \\ 0.5 or 50\%\end{tabular} \\ \hline
Quick ratio & \begin{tabular}[c]{@{}l@{}}The quick ratio is a primary liquidity ratio \\ that considers the difference between \\ immediate liquidity and deferred liquidity. 
\cite{liquidita}\end{tabular} & 
\begin{tabular}[c]{@{}l@{}} Greater than or \\ equal to 2, or less \\ than 2 but close to\end{tabular} & \begin{tabular}[c]{@{}l@{}} Less than or equal \\ to 1, or greater \\ than 1 but close \\ to \end{tabular}\\ \hline
\begin{tabular}[c]{@{}l@{}}Availability \\ index\end{tabular} & \begin{tabular}[c]{@{}l@{}}The availability index, which is also called \\ current ratio, is a secondary liquidity \\ index, therefore it offers a general indication \\ of the levels of liquidity in the company. \cite{liquidita}\end{tabular} & 
\begin{tabular}[c]{@{}l@{}} Greater than or \\ equal to 2, or less \\ than 2 but close to\end{tabular} & \begin{tabular}[c]{@{}l@{}} Less than or equal \\ to 1, or greater \\ than 1 but close \\ to \end{tabular}\\ \hline
\begin{tabular}[c]{@{}l@{}}Return of \\ capital\end{tabular} & \begin{tabular}[c]{@{}l@{}}Return of capital (ROC) is a payment that an \\ investor receives as a portion of their original \\ investment. \cite{roc}\end{tabular} & 
\begin{tabular}[c]{@{}l@{}}The higher the  \\ better (no limit)\end{tabular} & Closest to 0 \\ \hline
\begin{tabular}[c]{@{}l@{}}Rotation of \\ current \\ assets\end{tabular} & \begin{tabular}[c]{@{}l@{}}The index falls within the category of \\ indicators of rotation of invested capital, in \\ this case focusing attention on working \\ capital, with the particularity of considering \\ the value of production as a vehicle for \\ rotation or renewal, rather than sales revenues \\ (as is done usually in budget analyses). \cite{attivo}\end{tabular} & 
\begin{tabular}[c]{@{}l@{}}The higher the  \\ better (no limit)\end{tabular} & No alert values \\ \hline
\begin{tabular}[c]{@{}l@{}}Warehouse \\ turnover \\ index\end{tabular} & \begin{tabular}[c]{@{}l@{}}The warehouse turnover index represents a \\ fundamental parameter for companies, as \\ it helps to evaluate the efficiency of the \\ inventory management system and to \\ intervene accordingly with the most \\ appropriate adjustments. \cite{rotazione}\end{tabular} & 
\begin{tabular}[c]{@{}l@{}}The higher the  \\ better (no limit)\end{tabular} & No alert values \\ \hline
\begin{tabular}[c]{@{}l@{}}Solvency \\ index\end{tabular} & \begin{tabular}[c]{@{}l@{}}Solvency represents, in general, the ability \\ of a debtor to maintain the commitments \\ made with its creditors, repaying the money \\ received on loan in full and within the \\ appropriate timeframe. 
\cite{solvibilita}\end{tabular} & 
\begin{tabular}[c]{@{}l@{}}Solvency ratio \\ at least equal  \\ to 1 (greater \\ is excellent \\ degree of \\ solvency)\end{tabular} & \begin{tabular}[c]{@{}l@{}}Solvency \\ ratio less \\ than 1\end{tabular} \\ \hline
\end{tabular}
\end{table}

\begin{table}[t]
\centering
\caption{List of financial services}
\label{services}
\scriptsize
\begin{tabular}{|l|l|l|l|}
\hline
\begin{tabular}[c]{@{}l@{}}Financial \\ service name\end{tabular} & Description & Category & Indexes \\ \hline
\begin{tabular}[c]{@{}l@{}}Receivables \\ Discounting \\ o \\ Invoice \\ discounting\end{tabular} & \begin{tabular}[c]{@{}l@{}}The company assigns its trade receivables to a bank \\ or other financial institution, which advances a \\ portion of the value of the receivables, \\ withholding a fee for thefinancing service. \\ The bank or financial institution assumes the risk \\ of the debtor's insolvency and manages the \\ collection of the debt.\end{tabular} & \begin{tabular}[c]{@{}l@{}}Receivables \\ financing\end{tabular} & \begin{tabular}[c]{@{}l@{}}Liquidity \\ ratios\end{tabular} \\ \hline
Factoring & \begin{tabular}[c]{@{}l@{}}The company assigns its trade receivables to a \\ specialized company (Factor) which advances a \\ part of the value of the receivables, retaining a \\ commission for the financing and credit \\ management service. The Factor assumes the risk \\ of the debtor's insolvency and manages the \\ collection of the receivable.\end{tabular} & \begin{tabular}[c]{@{}l@{}}Receivables \\ financing\end{tabular} & \begin{tabular}[c]{@{}l@{}}Rotation of \\ current assets\end{tabular} \\ \hline
\begin{tabular}[c]{@{}l@{}}Account \\ Receivables \\ Securitization\end{tabular} & \begin{tabular}[c]{@{}l@{}}The company transfers its trade receivables to an  \\ external company (Securitiser) which uses them to \\ issue ABS which will later be bought by backers. \\
More on this service in the Securitization section \\ (see section \ref{Securitization}).
\end{tabular} & 
\begin{tabular}[c]{@{}l@{}}Receivables \\ financing\end{tabular} & \begin{tabular}[c]{@{}l@{}} Liquidity \\ ratios \end{tabular} \\ \hline
\begin{tabular}[c]{@{}l@{}}Dynamic \\ Discounting\end{tabular} & \begin{tabular}[c]{@{}l@{}}The supplier uploads the invoices to the dynamic \\ discounting portal, the customer approves the \\ invoices and proposes a discount in exchange for an \\ advance payment,the supplier accepts or rejects the \\ discount,the prepayment goes directly from the\\ customer to the supplier, the supplier receives 100\% \\ of the "discounted" amount.\end{tabular} & \begin{tabular}[c]{@{}l@{}}Financial \\ debts\end{tabular} & \begin{tabular}[c]{@{}l@{}}Return of \\ Capital\end{tabular} \\ \hline
\begin{tabular}[c]{@{}l@{}}Inventory \\ Financing \\ o Warehouse \\ receipt finance\end{tabular} & \begin{tabular}[c]{@{}l@{}}The company sells the goods in stock or in storage \\ to a bank or other financial institution, which \\ advances a portion of the value of the goods,\\ withholding a commission.\end{tabular} & \begin{tabular}[c]{@{}l@{}}Loan or \\ Advance \\ Financing\end{tabular} & \begin{tabular}[c]{@{}l@{}}Warehouse \\ turnover \\ index\end{tabular} \\ \hline
\end{tabular}
\end{table}

\section{Case study}
\label{case}

To define the objectives of the software platform, 
we present a simplified depiction of a typical supply chain that usually involves several key stages, production processes, stakeholders, and other actors.\\ 
The characteristic stages include the production of raw materials, transformation into finished goods, storage, and sale. The production process can encompass many product types, including consumer electronics, pharmaceuticals, and food. Stakeholders are individuals or organizations that participate in the activities of a supply chain and are influenced by the operations and outcomes that occur within it, such as suppliers, manufacturers, retailers, or final consumers.\\
Financial stability of stakeholders is crucial for the normal functioning of the supply chain. Liquidity risk represents a significant threat, and its mismanagement can result in substantial financial losses.  
Economic agreements (sales contracts) are often established between supply chain actors based on credits and debits, which are settled after a designated period. During this time, the creditor only holds a promise to pay, which can create problems if the debtor fails to comply, potentially disrupting the supply chain.
Furthermore, it is challenging to highlight the creditworthiness of a party in the absence of a financial institution overseeing the agreement. Without such monitoring, a stakeholder lacks the assurance that the debtor is a reliable counterpart.\\
The objective is to create an open and collaborative system that enables internal and external actors to share and observe the activities occurring within the supply chain, such as exchanging goods that result in debits and credits. The certification of financial operations in a distributed ledger, such as a blockchain, is a key aspect of this system. In this context, stakeholders and actors participate in consensus to ensure the integrity of the information. By sharing transactions and analyzing historical data, each stakeholder can manage their financial risks, including the possibility that a debtor will not pay. Exposing information to external financiers can attract new investments by transparently showing them the chain's health and allowing them to offer financing at more or less advantageous rates depending on their reputations.\\
Another objective is to implement real financial services by defining smart contracts to counter potential problems that may occur, such as the lack of liquidity, ensuring uninterrupted money flow. The service most closely examined for the proposed case study is credit securitization.

\subsection{Securitization}
\label{Securitization}
\begin{figure}[t]
\centering
\includegraphics[width=1.0\textwidth]{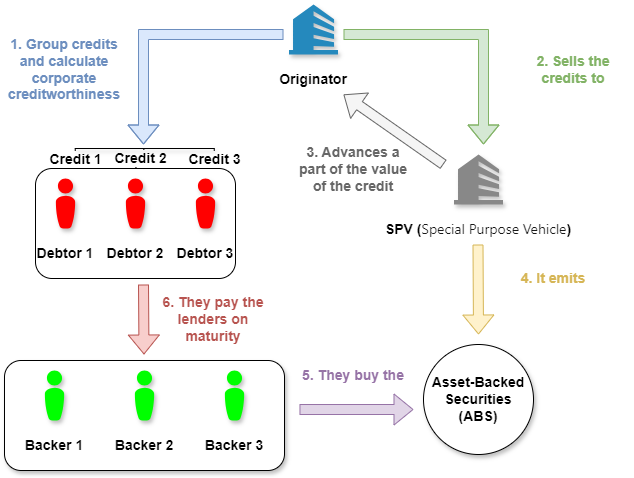}
\caption{Securitization schema} \label{cart}
\end{figure}

Securitization \cite{Ioannou} is a financial process in which a subject, for example, a bank or a company, called an originator, transfers a set of financial assets, such as loans or debts, to an SPV (Special Purpose Vehicle) for consideration. The latter advances a part of the value of the credit to the originator. Subsequently, the SPV issues ABS (Asset Backed Securities) to be resold on the market to cover the cost of purchasing the credits. In short, this process allows an entity to obtain immediate liquidity by transferring the risk of private default to the SPV and subsequently to the buyers of the ABS. Furthermore, the latter obtain risk diversification since they are not exposed to a single loan or credit but to a portfolio of assets. Figure \ref{cart} overviews the securitization process, illustrating the critical steps involved.

\section{The software platform}
\label{platform} 

\begin{figure}[t]
\centering
\includegraphics[width=1\textwidth]{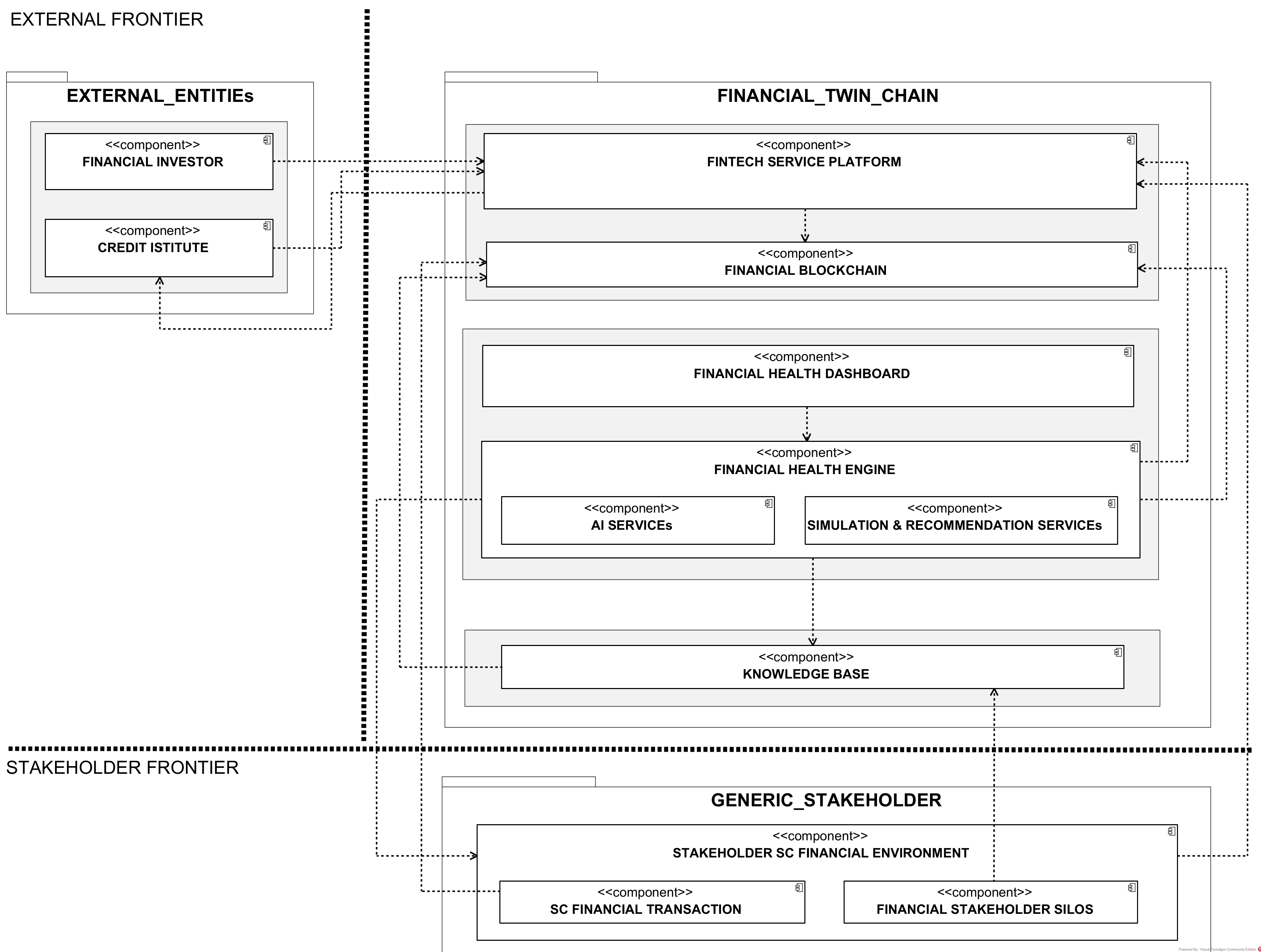}
\caption{Component Diagram (UML) of the software platform of the project.} \label{comdiag}
\end{figure}

A supply chain digital twin architecture has been designed to utilize the financial and non-financial data produced by various stakeholders. This architecture, shown in Figure \ref{comdiag}, collects and processes data to create a live twin representation of the entire supply chain, enabling continuous monitoring of its financial health and predicting financial issues. This will allow the suggestion of the utilization of a tailored subset of fintech services within the supply chain financial context.\\
The architectural description will commence with a bottom-up approach, illustrating the virtual twin construction as a harmonic big data flow that reaches a semantic improvement to analyze in real-time the sustainability of the entire value chain. This objective is of great importance and strategic value, and it can only be achieved by combining the big data flow, their semantic relations, the possible inferences, and the creation of wisdom. The output of this initial complex process generates a semantic big data layer, which is used as input for the subsequent step. This step operates to monitor the state of financial aspects of the chain, prevent problems through AI prediction functionalities, simulate operations in complex scenarios and identify the most appropriate subset of fintech solutions to apply for the specific sustainability instance of the chain. The fintech portal represents a significant component of the proposed design, it employs blockchain technology to facilitate the integration of pluggable SCF services, based on the stakeholders’ financial transactions.

\subsection{Stakeholder Frontier}
Stakeholder frontier is the source that produces two fundamental types of data:
\begin{itemize}
    \item \emph{SC Financial transaction}, derived by applying specific contracts that regulate the passage of an asset or service furniture between two or more actors named stakeholders. In our environment, the contracts are translated into smart contracts, and their operations are stored on the blockchain. This workflow contributes to any improvement in the financial supply chain achieved by using blockchain. This modality can augment the security, the trustability, and the visualization of the big mole of the economic data produced.
    \item \emph{Financial Stakeholder Silos}, are private stakeholders’ information, derived by internal stakeholders’ silos, that is voluntarily shared to augment the significant information used to achieve its actual grade of trustability and risk assessment.
\end{itemize}

\subsection{Financial Twin Chain}
The Financial Twin Chain is the system proposed in this paper, which ingests the big data produced by the Stakeholder Frontier and applies correlation and manipulation techniques to prevent financial waste. The system is composed of the following components:

\begin{itemize}
    \item \emph{Knowledge base module}, that uses data ingestion to consume the input and produce a semantic representation and correlation of financial and siloed input data, enabling the construction and population of a knowledge graph that can facilitate value creation through inference processes and the use of graph data science tools and applications. We can use all the derived semantic enrichment to apply predictive and monitoring logic. 

   \item \emph{Financial Health Engine}, that uses blockchain financial registrations and data from the knowledge base to apply artificial intelligence (AI) algorithms, such as predictive machine learning, to address financial sustainability issues. Furthermore, simulations and recommendations can be generated by constructing the virtual twin of the processes or assets obtained through appropriate semantic correlations. In summary, the Financial Health Engine component enables the continuous, real-time monitoring of financial key performance indicators (KPIs) or domain rules, such as the indexes and ratios presented in Table \ref{index}, predicts potential financial sustainability issues, generates simulations and recommendations, and suggests tailored fintech services to address the predicted sustainability problems.
    \item \emph{The Financial Health Dashboard}, employed to generate a range of detailed and related visualizations of the data generated as previously described. This component may be regarded as the optimal representation of a subset of information to achieve a specific visualization objective.  
    \item \emph{Financial blockchain}, the distributed ledger network in which stakeholders and actors participate in certifying financial transactions that occur throughout the lifecycle of a supply chain. They also collaborate to implement fintech services through the execution of smart contracts. Further details will be provided in a subsequent section.
    \item \emph{Fintech Service Platform}, a portal offering a wide range of fintech services to supply chain stakeholders and external actors. These services are provided using the financial blockchain and smart contracts. Further details will be provided in a subsequent section. 

\end{itemize}
\subsection{External Frontier}
The External Frontier encompasses financial investors and credit institutions that can contribute to the financial sustainability of the supply chain by investing as active participants in utilizing fintech services provided by the Fintech Services Platform of the Financial Twin Chain. This frontier includes all potential external actors interested in investing in the supply chain. Alternatively, to achieve optimal accessibility, standard fintech protocols (such as Open Banking \cite{open_banking}) can be employed.

\subsection{Financial blockchain}
We propose a blockchain-based ledger where stakeholders can store financial transactions that occur during the lifecycle of the supply chain. Each stakeholder participates in a permissioned network with their node, which maintains a copy of the ledger and contributes to the consensus mechanism. Stakeholders may:

\begin{itemize}
    \item Determine which actors and nodes can participate in the network and the extent of their permissions.
    \item Create, instantiate, and execute smart contracts that govern the financial transactions in the supply chain and implement the services offered by the fintech service platform.
    \item  Record new transactions to the ledger when authorized actors send requests to the network.
\end{itemize}
The purpose of the proposed blockchain is to facilitate the sharing of financial information, which is typically fragmented and accessible only to the parties directly involved among all stakeholders in a supply chain. This enables a clear picture of the state of the supply chain to be obtained, which allows potential critical issues to be identified and acted upon. The active involvement of stakeholders in the consensus process for the execution of operations and the guaranteed properties of non-modifiability, non-repudiation, and non-translatability in time of the ledger's transactions enhance the trust and transparency between stakeholders and improve cooperation between them. The involvement of external actors in the network, either as authorized users or as actual nodes participating in the consensus, can further increase the value of the chain by providing certified evidence of its robustness, with the potential to attract new investors and offer new services by engaging financial institutions (such as a bank).

\subsection{Fintech service platform}
\begin{figure}
\centering
\includegraphics[width=1\textwidth]{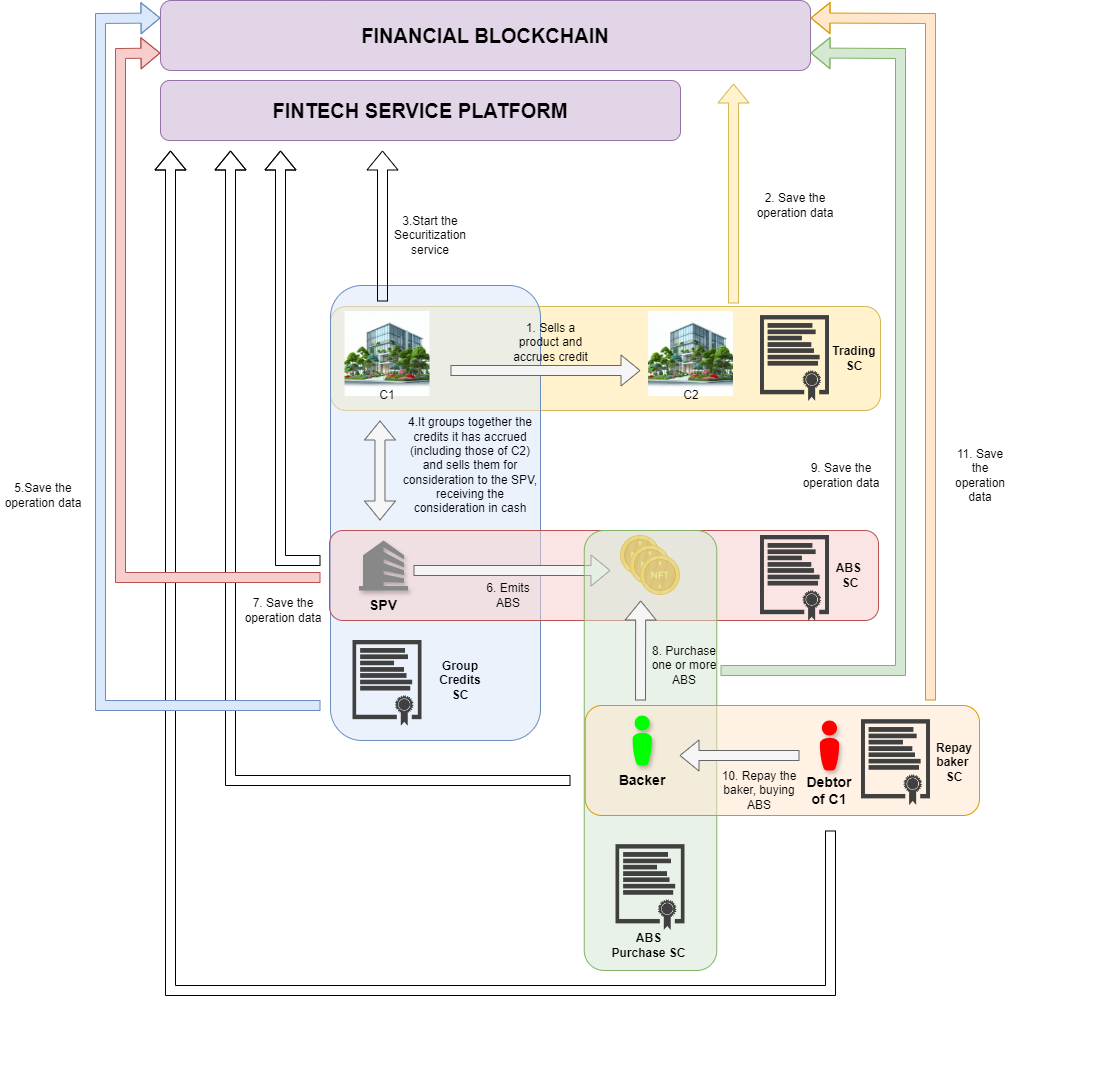}
\caption{Use of smart contracts and Fintech service platform to implement securitization} \label{ex_cart}
\end{figure}

The Fintech service platform is a digital portal that offers a wide range of financial tools and services within the supply chain. This portal serves as a hub where companies can access a variety of financial solutions, which are intended to optimize their business operations, improve cash flow management, and increase overall efficiency within the supply chain ecosystem. To manage this domain, it is important to include access to various financing options such as loans, factoring, or trade financing, which can help companies manage working capital needs and bridge gaps in liquidity or finance operations.\\
The proposed platform would facilitate the implementation of a transparent and automated credit securitization service by utilizing the financial blockchain and implementing dedicated smart contracts. To address potential liquidity issues, a stakeholder may wish to securitize one or more matured credits in the supply chain to obtain immediate liquidity. The information recorded in the proposed distributed ledger enables the definition of the financial behavior of the various actors involved, thus facilitating the implementation of smart contracts that regulate the sale of credits to SPVs. These contracts also associate financial indices concerning the debtor companies and the subsequent issue of ABS that can be purchased by interested backers.  Furthermore, the provided services will enable the debtors of the original credits to settle their debts on the stipulated date by paying the requisite amount directly to the purchasers of the bonds. Figure \ref{ex_cart} provides a schematic overview of all blockchain securitization transactions. The financial risk of the various transactions is calculated automatically via smart contracts and can be easily verified by all relevant actors. These parties have access to all the necessary information and tools to determine whether to proceed with the transactions or to agree on a purchase price that they deem appropriate based on the inherent risk. Banking institutions can facilitate payments directly in fiat currency or cryptocurrency implemented by the platform via fungible tokens (FT), which can be converted into real currency later.

\section{Conclusions}
\label{end}
The study proposes an architectural solution to address and mitigate financial criticality in a supply chain. In order to achieve this objective, it incorporates software components that collaborate to gather financial data from stakeholders, implement predictive models of financial risks as proposed in the literature, utilise blockchain technology to enhance trust and security over data, and develop financial services to optimise financial liquidity such as, for example, securitization. \\
The pluggable architectural design permits the addition of any machine learning model and data source from the supply chain operations domain for the purpose of financial risk prediction. Furthermore, the platform offers tools that facilitate the implementation of new measures and services, leveraging the potential of blockchain technology to enhance the financial health of the company.

\end{document}